\documentstyle[]{article}

\input{epsf}

\renewcommand{\abstract}[1]{{ \footnotesize \noindent {\bf Abstract} #1 \\}}
\renewcommand{\author}[1]{\subsubsection*{\bf#1}}
\newcommand{\address}[1]{\subsubsection*{\it#1}}

\begin{document}

\section*{\LARGE{\bf{Mass modelling of galaxy clusters via\\ velocity moments}}}

\vspace{0.3in}

\author{E. L. {\L}okas$^{1}$, R. Wojtak$^{1}$, G. A. Mamon$^{2}$ and S. Gottl\"ober$^{3}$}

\address{$^1$Nicolaus Copernicus Astronomical Center, Bartycka 18, 00-716 Warsaw, Poland\\
$^2$Institut d'Astrophysique de Paris, 98 bis Bd Arago, F-75014 Paris, France\\
$^3$Astrophysikalisches Institut Potsdam, An der Sternwarte 16, 14482 Potsdam, Germany}

\abstract{We summarize the method of mass modelling of galaxy clusters based on
reproducing the dispersion and kurtosis of the projected velocity distribution
of galaxies. The models are parametrized within the framework of the NFW density profile,
characterized by the virial mass and concentration, together with the constant anisotropy
of galaxy orbits. The use of velocity dispersion alone does not allow to constrain
all the three parameters from kinematic data due to the mass-anisotropy degeneracy. The
degeneracy is broken by introducing the fourth velocity moment, the kurtosis. We tested the
method based on fitting both moments on mock data sets drawn from simulated dark matter haloes
and showed it to reproduce reliably the properties of the haloes. The method has been applied
to estimate the mass, concentration and anisotropy of more than 20 clusters which allowed us to
confirm, for the first time using kinematic data, the mass-concentration relation
found in N-body simulations.}

\section{The method of velocity moments}

The traditional approach to studying the kinematic data for galaxy clusters involves the modelling
of the velocity dispersion profile. The procedure consists of collecting enough (preferably more than
a hundred) redshifts and positions of the galaxies, binning the data and calculating the dispersion
values as a function of distance from the cluster centre.
In the parametric approach one then assumes that the mass distribution is given by some model,
e.g. the NFW (Navarro, Frenk \& White 1997) profile characterized by the virial mass $M_v$
and concentration $c$,
and makes some assumption about the anisotropy of galaxy orbits $\beta$ (we assume $\beta$=const
with radius). The modelling then involves solving
the lowest-order Jeans equation for spherical systems (Binney \& Tremaine 1987), projecting the
solutions along the line of sight and adjusting the free parameters.

\begin{figure}
\begin{center}
    \leavevmode
    \epsfxsize=6cm
    \epsfbox[90 15 290 205]{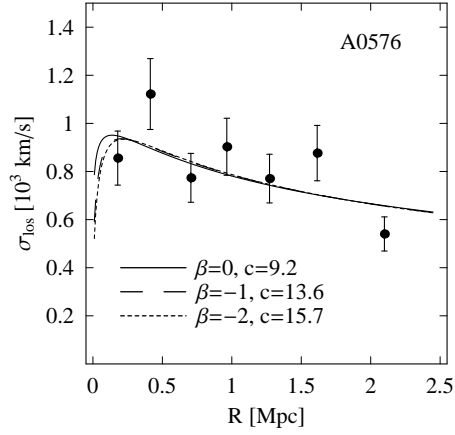}
\end{center}
\caption{The velocity dispersion profile of the cluster Abell 576. The data points show the dispersions
calculated with 30 galaxies per bin with standard sampling errors. The three lines show three examples of
projected solutions of the Jeans equation with indicated parameters. }
\label{a576_dispersion}
\end{figure}

An example of such modelling for the case of Abell 576 is shown in Figure~\ref{a576_dispersion}.
As we can see, for any anisotropy parameter $\beta$
a value of concentration $c$ of the density profile can be found which provides a good fit, which signifies
the degeneracy between the parameters. Both parameters cannot be constrained from the dispersion alone.
One can break this degeneracy by considering the fourth-order velocity moment, the kurtosis, which depends
mainly on anisotropy (see Sanchis et al. 2004). The improved method thus involves solving two Jeans
equations for the radial moments (see {\L}okas 2002)
\begin{eqnarray}
        \frac{\rm d}{{\rm d} r}  (\nu \sigma_r^2) + \frac{2 \beta}{r} \nu
	\sigma_r^2 + \nu \frac{{\rm d} \Phi}{{\rm d} r} &=& 0 	\label{jeans_equation1}  \\
        \frac{\rm d}{{\rm d} r}  (\nu \overline{v_r^4}) + \frac{2 \beta}{r} \nu
	\overline{v_r^4} + 3 \nu \sigma_r^2 \frac{{\rm d} \Phi}{{\rm d} r} &=& 0    \label{jeans_equation2}
\end{eqnarray}
and projecting them along the line of sight
\begin{eqnarray}
	\sigma_{\rm los}^2 (R) &=& \frac{2}{I(R)} \int_{R}^{\infty}
	\frac{\nu \sigma_r^2 r}{\sqrt{r^2 - R^2}} \left( 1-\beta \frac{R^2}{r^2} \right)
	{\rm d} r                        \label{proj_solution1}  \\
	\overline{v_{\rm los}^4} (R) &=& \frac{2}{I(R)} \int_{R}^{\infty}
	\frac{\nu \,  \overline{v_r^4} \,r}{\sqrt{r^2 - R^2}} \ \left[ 1 - 2 \beta \frac{R^2}{r^2}
	+ \beta(1+\beta) \frac{R^4}{2 r^4} \right]
	\,{\rm d} r   \label{proj_solution2}
\end{eqnarray}
where $\nu$ is the distribution
of the tracer (which we assume to follow the mass), $I(R)$ is its 2D projection and $\Phi$ is the gravitational
potential. It is convenient to express the fourth moment in the form of the line-of-sight or
projected kurtosis $\kappa_{\rm los} (R) = \overline{v_{\rm los}^4} (R)/\sigma_{\rm los}^4 (R)$
whose value is 3 for a Gaussian distribution.

\section{Tests on N-body simulations}

\begin{figure}
\begin{center}
    \leavevmode
    \epsfxsize=12cm
    \epsfbox[100 30 570 180]{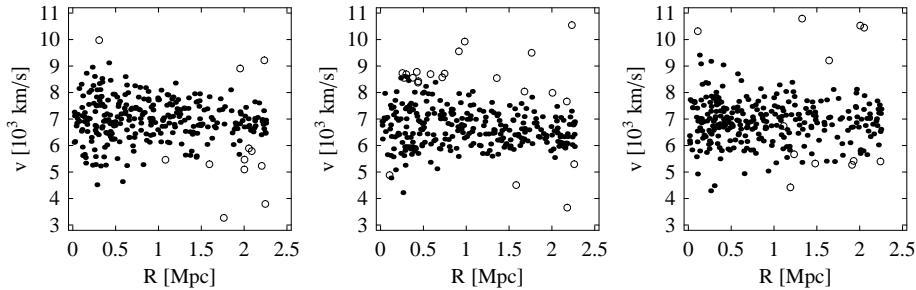}
\end{center}
\caption{Mock kinematic data sets generated from the simulated dark matter haloes.}
\label{mock_data}
\end{figure}

The method has been tested on mock data samples generated from simulated dark matter haloes (for details
see Sanchis et al. 2004 and {\L}okas et al. 2006). We have used a set of ten dark matter haloes formed
in a standard $\Lambda$CDM cosmology described by Wojtak et al. (2005). The mock data sets were constructed
by placing an imaginary observer at a distance of 100 Mpc from a given halo and projecting all particle
velocities along the observer's line of sight (for simplicity: along one of the axes of the simulation box)
and their positions on the plane of the sky. We then
randomly selected 300 particles from the projected phase space of each halo limiting the particle velocities to the
range $\pm4000$ km s$^{-1}$ with respect to the halo mean velocity.

\begin{figure}
\begin{center}
    \leavevmode
    \epsfxsize=12cm
    \epsfbox[100 35 580 180]{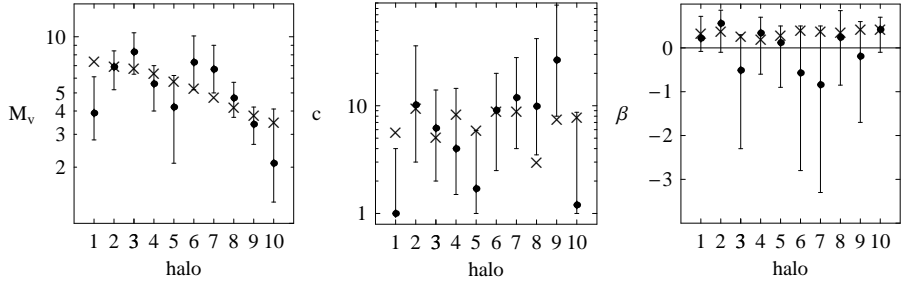}
\end{center}
\caption{Comparison of the fitted parameters (dots with error bars) with the true parameter values measured
from the simulation (crosses). The virial masses are in units of $10^{14}$ M$_{\odot}$.}
\label{fitted_haloes}
\end{figure}

Three examples of such mock data sets are shown in Figure~\ref{mock_data} for one of the haloes
(halo 2 in Wojtak et al. 2005 and {\L}okas et al. 2006)
with the line of sight along the $x,y$ and $z$ axis of the simulation box. Open symbols mark the particles
unbound to the cluster but present in the samples due to projection effects. Such contamination has to be
removed before the proper modelling at least in the case of the strongest outliers which would otherwise
artificially inflate the values of velocity moments and bias the estimated parameters of the cluster. This can
be done by a number of methods (for a review see Wojtak et al. 2007). Once the samples are cleaned of
interlopers we calculate the profiles of velocity moments and fit them with solutions of the Jeans
equations (\ref{proj_solution1})-(\ref{proj_solution2}) estimating the best-fitting parameters $M_v$, $c$
and $\beta$. The results of such tests are presented in Figure~\ref{fitted_haloes} for all ten haloes
observed along the $y$ axis of the simulation box. The recovered parameter values in most cases
agree within errors with the true values measured from the full 3D information about the haloes.

\section{Application to nearby clusters}

\begin{figure}[h]
\begin{center}
    \leavevmode
    \epsfxsize=12cm
    \epsfbox[105 30 580 180]{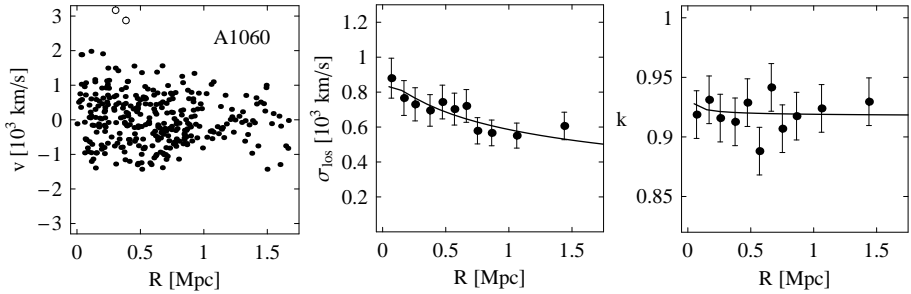}
\end{center}
\caption{The kinematic sample used for the modelling of A1060 (left), the velocity dispersion profile
(middle) and the kurtosis profile (right). The lines show the best-fitting solutions of the Jeans
equations.}
\label{a1060all}
\end{figure}

The method of velocity moments as described above has been applied up till now to more than 20 nearby
($z<0.1$) rich galaxy clusters starting with the Coma cluster ({\L}okas \& Mamon 2003), and following-up
with other nearby relaxed clusters ({\L}okas et al. 2006; Wojtak \& {\L}okas 2007). An example of the
results for the cluster Abell 1060 is presented in Figure~\ref{a1060all}. The left panel shows the
kinematic sample of 330 member galaxies used for the modelling with two open symbols indicating the
rejected outliers. In the middle panel we plot the velocity dispersion profile, which as expected for
systems with NFW-like density profile and isotropic orbits ({\L}okas \& Mamon 2001), decreases slowly
with distance.
The right panel plots the kurtosis-like variable $k=(\log \kappa)^{1/10}$ which has
the advantage of a Gaussian sampling distribution (see {\L}okas \& Mamon 2003). In both panels the solid
lines show the best-fitting solutions of the Jeans equations (\ref{proj_solution1})-(\ref{proj_solution2})
obtained for $M_v=4.4^{+1.1}_{-1.0} \times 10^{14}$ M$_{\odot}$, $c=14^{+22}_{-10}$ and
$\beta=0.03^{+0.72}_{-1.13}$, where the errors are 1$\sigma$
uncertainties following from the sampling errors of velocity moments.

\begin{figure}
\begin{center}
    \leavevmode
    \epsfxsize=6cm
    \epsfbox[90 10 290 210]{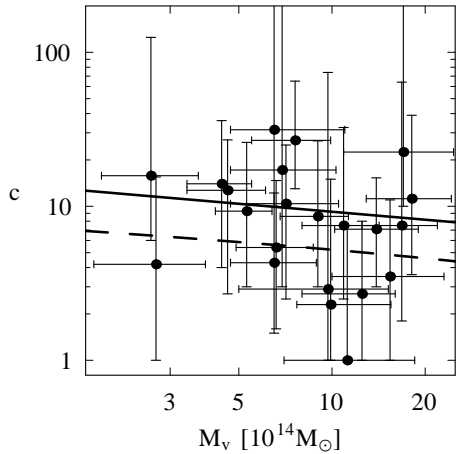}
\end{center}
\caption{The mass-concentration relation from kinematic data for 22 clusters. The solid line is the fit
to the data points, the dashed line is the prediction from the simulations of Bullock et al. (2001).}
\label{cm22clusters}
\end{figure}

Similar modelling for other clusters has been performed giving us the possibility to verify the
mass-concentration relation postulated e.g. by NFW and Bullock et al. (2001). The relation obtained
by combining data for 22 clusters is shown in Figure~\ref{cm22clusters}. The fit to the data points
of the relation $\log c = a \log [M_v/(10^{14} M_{\odot})] + b $ gives
$a=-0.17 \pm 0.65$ and $b=1.14 \pm 0.62$ (1$\sigma$ error bars). This fit is plotted as a solid line
in Figure~\ref{cm22clusters}. We can see that the slope of the relation agrees well with the
prediction from N-body simulations of dark matter haloes by Bullock et al. (2001) shown with the dashed line,
although the concentrations from the data are typically shifted towards higher values for a given mass.

\end{document}